\documentclass[preprint,eqsecnum,aps,nofootinbib]{revtex4}
\usepackage{amsfonts,amsmath,amssymb,amsthm}
\usepackage{latexsym}
\usepackage{bbm,bm}
\usepackage{graphicx}

\newcommand{\ket}[1]{\lvert #1 \rangle}
\newcommand{\bra}[1]{\langle #1 \lvert}
\newcommand{\beq}{\begin{equation}}
\newcommand{\eeq}{\end{equation}}
\newcommand{\beqs}{\begin{eqnarray}}
\newcommand{\eeqs}{\end{eqnarray}}

\begin{document}

\title{REE From EOF}

\author{Eylee Jung$^{1}$ and DaeKil Park$^{1,2}$\footnote{dkpark@kyungnam.ac.kr}}

\affiliation{$^1$Department of Electronic Engineering, Kyungnam University, Changwon
                 631-701, Korea    \\
             $^2$Department of Physics, Kyungnam University, Changwon
                  631-701, Korea    
                      }

\begin{abstract}
It is well-known that entanglement of formation (EOF) and relative entropy of entanglement (REE)
are exactly identical for all two-qubit pure states even though their definitions are completely
different. We think this fact implies that there is a veiled connection between EOF and REE.
In this context, we suggest a procedure, which enables us to compute REE from EOF without relying
on the converse procedure. It is shown that the procedure yields correct REE for many symmetric mixed states 
such as Bell-diagonal, generalized Vedral-Plenino, and generalized Horodecki states. It also gives a
correct REE for less symmetric Vedral-Plenio-type state. However, it is shown that the procedure 
does not provide correct REE for arbitrary mixed states. 
\end{abstract}

\maketitle
\section{Introduction}
Entanglement of formation (EOF)\cite{benn96} and relative entropy of entanglement (REE)\cite{vedral-97-1,vedral-97-2} are two major 
entanglement monotones for bipartite systems. For pure states $\rho = \ket{\psi} \bra{\psi}$
the EOF ${\cal E}_F (\rho)$ is defined as a von Neumann entropy of its subsystem $\rho_A = \mbox{tr}_B \rho$.
On the contrary,
REE is defined as minimum value of the relative entropy  with separable states;
\begin{equation}
\label{ree-1-1}
{\cal E}_R (\rho) = \min_{\sigma \in {\cal D}} \mbox{tr} (\rho \ln \rho - \rho \ln \sigma),
\end{equation}
where ${\cal D}$ is a set of separable states\footnote{Since REE is defined through another separable state $\sigma$, it is called 
``distance entanglement measure''. Another example of the distance entanglement measure is a geometric entanglement measure defined as $E_g (\psi)= 1-P_{max}$, 
where $P_{max}$ is a maximal overlap of a given state $\ket{\psi}$ with the nearest product state\cite{groverian}.}.  
It was shown in Ref.\cite{vedral-97-2} that $E_{R} (\rho)$ 
is a upper bound of the distillable entanglement\cite{benn96}. The separable state $\sigma_*$, which yields a minimum value of the relative entropy
is called the closest separable state (CSS) of $\rho$. Surprising fact, at least for us, is that although definitions of EOF and REE
are completely different, they are exactly same for all pure states\cite{vedral-97-2}. This fact may indicate that they are related to each other 
although the exact connection is not revealed yet. The main purpose of this paper is to explore the veiled connection between EOF and REE.
 
For mixed states $\rho$ EOF is defined via a convex-roof method\cite{benn96,uhlmann99-1};
\begin{equation}
\label{two3}
{\cal E}_F (\rho) = \min \sum_i p_{i} {\cal E}_F (\rho_i),
\end{equation}
where the minimum is taken over all possible pure-state decompositions with $0 \leq p_i \leq 1$ and $\sum_i p_i = 1$. The ensemble that gives 
the minimum value in Eq.(\ref{two3}) is called the optimal decomposition of the mixed state $\rho$.
Thus, the main task for analytic calculation of EOF is derivation of an optimal decomposition of the given mixture. Few years ago, 
the procedure for construction of the optimal decomposition was derived\cite{hill97,woot-98} in 
the two-qubit system, the simplest bipartite system, 
by making use of the time-reversal operation of spin-1/2 particles appropriately. In these references the relation
\begin{equation}
\label{EoF-1}
{\cal E}_F (C) = h \left( \frac{1 + \sqrt{1 - C^2}}{2} \right)
\end{equation}
is used, where $h(x)$ is a binary entropy function $h(x) = -x \ln x - (1 - x) \ln (1 - x)$ and $C$ is called the concurrence.
This procedure, usually called Wootters procedure, was re-examined in Ref.\cite{uhlmann99-1} in terms of antilinearity.
Introduction of antilinearity in quantum information theory makes it possible to derive concurrence-based entanglement monotones for 
tripartite\cite{ckw} and multipartite systems\cite{concurrence-b}. Due to the discovery of the closed formula
for EOF in the two-qubit system, EOF is recently applied not only 
to quantum information theory but also to many scientific fields such as life science\cite{life}.

While EOF is used in various areas of science, REE is not because of its calculational difficulty. In order to obtain REE analytically 
for given mixed state $\rho$ one should derive its CSS, but still we don't know how to derive CSS\cite{open} even in the two-qubit system except 
very rare cases\cite{vedral-97-2,difficulty,z-axis}. In Ref.\cite{difficulty} REE for Bell-diagonal, generalized Vedral-Plenio\cite{vedral-97-2}, 
and generalized Horodecki states\cite{horodeckis} were derived analytically through pure geometric arguments\cite{horodecki96}. 

Due to the notorious difficulty some people try to solve the REE problem conversely. Let $\sigma_*$ be a two-qubit boundary states in 
the convex set of the separable states. In Ref.\cite{miran-08-1} authors derived entangled states, whose CSS are $\sigma_*$. This 
converse procedure is extended to the qudit system\cite{converse-1} and is generalized as convex optimization problems\cite{converse-2}.
However, as emphasized in Ref.\cite{difficulty} still it is difficult to find a CSS $\sigma_*$ of given entangled state $\rho$ although
the converse procedure may provide some useful information on the CSS\cite{z-axis}.

In this paper we will try to find a CSS for given entangled two-qubit state without relying on the converse procedure. As commented, EOF and REE are identical for bipartite pure
states although they are defined differently. This means that they are somehow related to each other. If this connection is unveiled, probably
we can find CSS for arbitrary two-qubit mixed states because we already know how to compute EOF through Wootters procedure. To explore this issue is 
original motivation of this paper. We will show in the following that REE of many mixed symmetric states can be analytically obtained from EOF if 
one follows the following procedure: 
\begin{enumerate}
  \item[{\em (1)}] 
        For entangled two-qubit state $\rho$ let $\rho = \sum_j p_j \rho_j \hspace{.3cm} (\rho_j = \ket{\psi_j} \bra{\psi_j})$ 
        be an optimal decomposition for calculation of EOF. 
  \item [{\em (2)}] 
        Since $\rho_j$ are pure states, it is possible to obtain their CSS $\sigma_j$. Thus, it is straight to derive a separable mixture
        $\tilde{\sigma} = \sum_j p_j \sigma_j$.
  \item [{\em (3)}] 
        If $\tilde{\sigma}$ is a boundary state in the convex set of separable states, the procedure is terminated with $\sigma_* = \tilde{\sigma}$.
  \item [{\em (4)}]
        If $\tilde{\sigma}$ is not a boundary state, we consider $\pi = q \rho + (1-q) \tilde{\sigma}$. By requiring that $\pi$ is a boundary
        state, one can fix $q$, {\it say} $q=q_0$. Then we identify $\sigma_* = q_0 \rho + (1-q_0) \tilde{\sigma}$.
\end{enumerate}
This procedure is schematically represented in Fig. 1. 

In order to examine the validity of the procedure we have to apply the procedure to the mixed states whose REE are already known. Thus, we will choose the Bell-diagonal,
generalized Vedral-Plenio and generalized Horodecki states, whose REE were computed in Ref.\cite{vedral-97-2,difficulty,miran-08-1} through different methods. Also, we will 
apply the procedure to the less symmetric mixed states such as Vedral-Plenio-type and Horodecki-type states whose REE were computed in Ref.\cite{z-axis} by making use of the 
the converse procedure introduced in Ref.\cite{miran-08-1}.

The paper is organized as follows. 
In section II we show that the 
procedure generates the correct CSS for Bell-diagonal states. In section III and section IV we show that 
the procedure generates the correct CSS for generalized Vedral-Plenio and generalized Horodecki states, 
respectively. In section V we consider two less symmetric states, Vedral-Plenio-type and Horodecki-type 
states. It is shown that while the procedure generates a correct CSS for the former, it does not give a 
correct one for the latter. In section VI a brief conclusion is given.
In appendix we prove that EOF and REE are identical for all pure states by making use of the Schmidt decomposition. The Schmidt bases derived in this appendix are used in the 
main body of this paper.

\begin{figure}[ht!]
\begin{center}
\includegraphics[height=7cm]{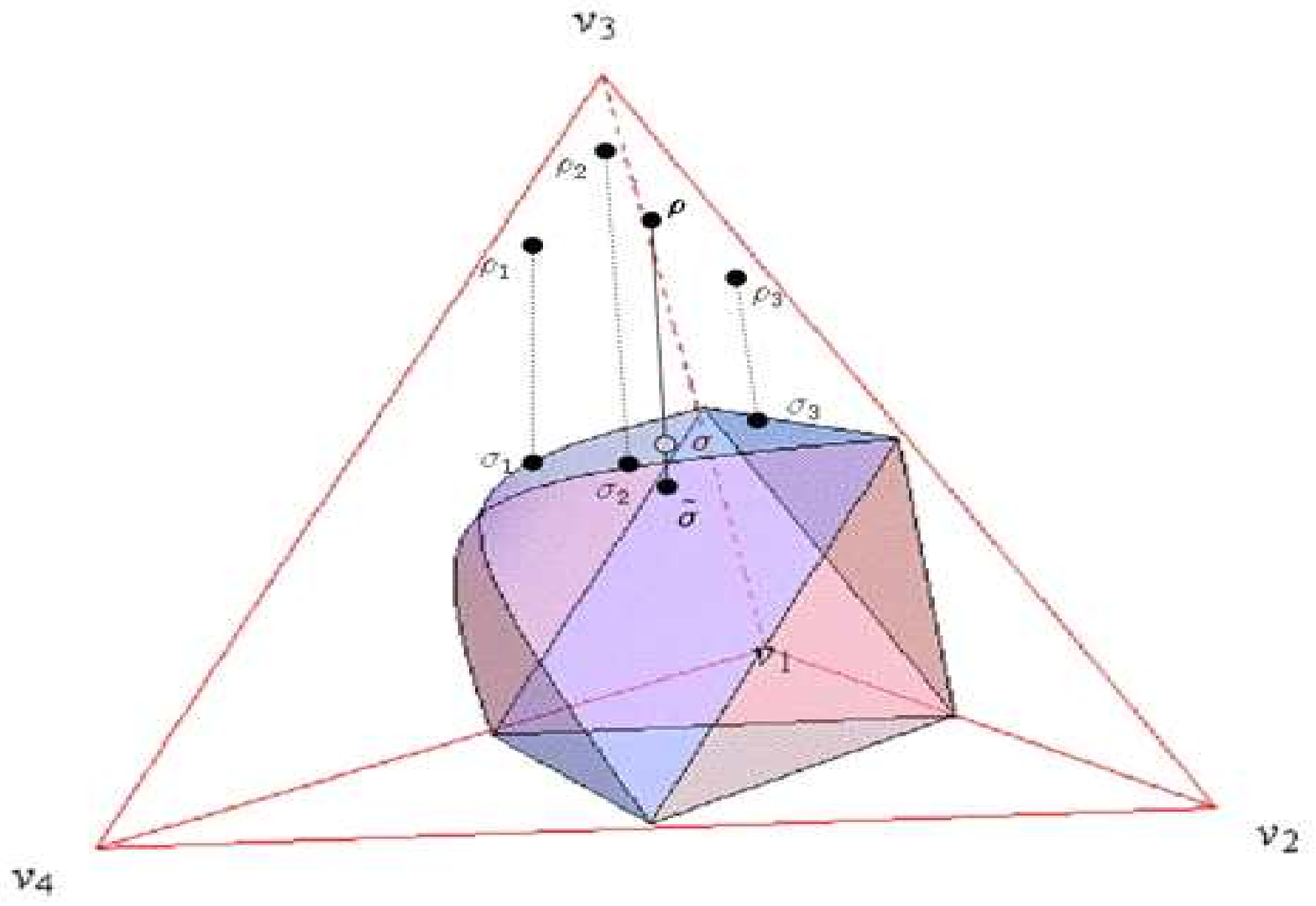}
\caption[fig1]{(Color online) The schematic diagram of the procedure, by which REE can be computed from EOF.
The 
polygon at the center is a deformed octahedron\cite{horodecki96,difficulty}. Inside and outside of the
octahedron separable and entangled states reside, respectively. The CSS of the entangled state resides at the 
surface of the octahedron.}
\end{center}
\end{figure}

\section{Bell-Diagonal States}
In this section we will show that the procedure mentioned above solves the REE problem of the Bell-diagonal states:
\begin{equation}
\label{BD-1}
\rho_{BD} = \lambda_1 \ket{\beta_1} \bra{\beta_1} + \lambda_2 \ket{\beta_2} \bra{\beta_2} + \lambda_3 \ket{\beta_3} \bra{\beta_3} + 
            \lambda_4 \ket{\beta_4} \bra{\beta_4}
\end{equation}
where $\sum_{j=1}^4 \lambda_j = 1$, and 
\begin{eqnarray}
\label{BD-2}
& &\ket{\beta_1} = \frac{1}{\sqrt{2}} (\ket{00} + \ket{11})  \hspace{1.0cm} \ket{\beta_2} = \frac{1}{\sqrt{2}} (\ket{00} - \ket{11})
                                                                                                                               \\   \nonumber
& &\ket{\beta_3} = \frac{1}{\sqrt{2}} (\ket{01} + \ket{10})  \hspace{1.0cm}  \ket{\beta_4} = \frac{1}{\sqrt{2}} (\ket{01} - \ket{10}).
\end{eqnarray}
The CSS and REE of $\rho_{BD}$ were obtained in many literatures\cite{vedral-97-2,difficulty,z-axis} through various different methods. 
If, for convenience, $\max(\lambda_1, \lambda_2, \lambda_3, \lambda_4) = \lambda_3$ , the CSS and REE of $\rho_{BD}$ are 
\begin{eqnarray}
\label{BD-3}
& &\pi_{BD} = \frac{\lambda_1}{2(1 - \lambda_3)} \ket{\beta_1}\bra{\beta_1} + 
           \frac{\lambda_2}{2(1 - \lambda_3)} \ket{\beta_2}\bra{\beta_2} + \frac{1}{2} \ket{\beta_3}\bra{\beta_3} + 
		   \frac{\lambda_4}{2(1 - \lambda_3)} \ket{\beta_4}\bra{\beta_4}                                \\   \nonumber
& &\hspace{3.0cm}    {\cal E}_R (\rho_{BD}) =  -h (\lambda_3) + \ln 2.
\end{eqnarray}

Now, we will show that the procedure we suggested also yields the same result.
Following Wootters procedure, one can show that the optimal decomposition of $\rho_{BD}$ for $\lambda_3 \geq 1/2$\footnote{If $\lambda_3 \leq 1/2$, $\rho_{BD}$ is a separable state.} is 
\begin{equation}
\label{BD-4}
\rho_{BD} = \sum_{j=0}^4 p_{j} \ket{\psi^{BD}_j} \bra{\psi^{BD}_j}
\end{equation}
where $p_j = 1/4 \hspace{.5cm} (j=1, \cdots, 4)$ and 
\begin{eqnarray}
\label{BD-5}
& &\ket{\psi^{BD}_1} = \sqrt{\lambda_1} \ket{\beta_1} + i \sqrt{\lambda_2} \ket{\beta_2} + \sqrt{\lambda_3} \ket{\beta_3} + 
                   \sqrt{\lambda_4} \ket{\beta_4}                                           \nonumber  \\
& &\ket{\psi^{BD}_2} = \sqrt{\lambda_1} \ket{\beta_1} + i \sqrt{\lambda_2} \ket{\beta_2} - \sqrt{\lambda_3} \ket{\beta_3} - 
                   \sqrt{\lambda_4} \ket{\beta_4}                                            \\   \nonumber
& &\ket{\psi^{BD}_3} = \sqrt{\lambda_1} \ket{\beta_1} - i \sqrt{\lambda_2} \ket{\beta_2} + \sqrt{\lambda_3} \ket{\beta_3} - 
                   \sqrt{\lambda_4} \ket{\beta_4}                                             \\   \nonumber
& &\ket{\psi^{BD}_4} = \sqrt{\lambda_1} \ket{\beta_1} - i \sqrt{\lambda_2} \ket{\beta_2} - \sqrt{\lambda_3} \ket{\beta_3} + 
                   \sqrt{\lambda_4} \ket{\beta_4}.
\end{eqnarray} 
All $\ket{\psi^{BD}_j} \hspace{.3cm} (j = 1, \cdots, 4)$ have the same concurrence 
${\cal C} = 2 \lambda_3 - 1$ and, hence, the same $\lambda_{\pm}$ (defined in Eq. (\ref{def-2-2})) as
\begin{equation}
\label{BD-6}
\lambda_{\pm} = \frac{1}{2} \left( \sqrt{\lambda_3} \pm \sqrt{1 - \lambda_3} \right)^2.
\end{equation}

The Schmidt bases of $\ket{\psi^{BD}_1}$ can be explicitly derived by following the procedure of appendix A and the result is 
\begin{eqnarray}
\label{BD-7}
& &\ket{0_A}                                                                      
= \frac{1}{N_+}
               \left[ \left(\sqrt{1 - \lambda_3} + \sqrt{\lambda_4}\right) \ket{0} + \left(\sqrt{\lambda_1} - i \sqrt{\lambda_2}\right)
                                                              \ket{1}  \right]  
                                                                                 \nonumber  \\
& &\ket{1_A}                                                                      
= \frac{-1}{N_-}
       \left[ \left(\sqrt{1 - \lambda_3} - \sqrt{\lambda_4}\right) \ket{0} - \left(\sqrt{\lambda_1} - i \sqrt{\lambda_2}\right)
                                                              \ket{1}  \right]    \\   \nonumber
& &\ket{0_B}                                                                      
= \frac{1}{N_+} \left[ \left(\sqrt{\lambda_1} + i \sqrt{\lambda_2}\right) \ket{0} + 
                  \left(\sqrt{1 - \lambda_3} + \sqrt{\lambda_4}\right)
                                                                \ket{1}  \right]  
                                                                                 \nonumber  \\
& &\ket{1_B}                                                                      
= \frac{1}{N_-}
      \left[ \left(\sqrt{\lambda_1} + i \sqrt{\lambda_2}\right) \ket{0} - 
                  \left(\sqrt{1 - \lambda_3} - \sqrt{\lambda_4}\right)
                                                                \ket{1}  \right],  
                                                                                 \nonumber
\end{eqnarray}
where the normalization constants $N_{\pm}$ are
\begin{equation}
\label{BD-8}
N_{\pm} =  \sqrt{2 \sqrt{1 - \lambda_3} \left(\sqrt{1 - \lambda_3} \pm \sqrt{\lambda_4}\right)}.
\end{equation}

Thus the CSS of $\ket{\psi^{BD}_1}$, say $\sigma_1$, can be straightforwardly computed by making 
use of Eq. (\ref{result-2});
\begin{eqnarray}
\label{BD-9}
\sigma_1 &=&  \lambda_+  \ket{0_A 0_B} \bra{0_A 0_B} + \lambda_- \ket{1_A 1_B} \bra{1_A 1_B} 
                                                                         \nonumber   \\
&=&  \frac{1}{4 \left(1 - \lambda_3 \right)}
\left(               \begin{array}{cccc}
                   \mu \mu^*& \mu \nu_+ & \mu \nu_- & \mu^2                           \\
                   \mu^* \nu_+  &  d_+  &  \mu \mu^*  &  \mu \nu_+                     \\
                   \mu^* \nu_-  &  \mu \mu^*  &  d_-  &  \mu \nu_-                    \\
                   \left(\mu^*\right)^2  &  \mu^* \nu_+  &  \mu^* \nu_-  &  \mu \mu^*
                       \end{array}                       \right),
\end{eqnarray}
where
\begin{eqnarray}
\label{BD-10}
& &\mu = \sqrt{\lambda_1} + i \sqrt{\lambda_2}  \hspace{1.0cm}
\nu_{\pm} = 2 \left( 1 - \lambda_3 \right) \sqrt{\lambda_3} \pm \sqrt{\lambda_4}   \\    \nonumber
& & \hspace{1.0cm} d_{\pm} = \left(1 - \lambda_3 + \lambda_4 \right) \pm 4 \left(1 - \lambda_3 \right)
                                              \sqrt{\lambda_3 \lambda_4}.
\end{eqnarray}
Similarly, one can derive the Schmidt bases for other  $\ket{\psi^{BD}_j} \hspace{.2cm} (j=2,3,4)$
and the corresponding CSS $\sigma_j$. Then, one can show that the separable state 
$\tilde{\sigma} = \sum_{j=1}^4 p_j \sigma_j$ with $p_j = 1/4$ for all $j$ is 
\begin{eqnarray}
\label{BD-11}
\tilde{\sigma} = \frac{1}{4 \left(1 - \lambda_3 \right)}
\left(               \begin{array}{cccc}
     \lambda_1 + \lambda_2  &  0  &  0  &  \lambda_1 - \lambda_2    \\
     0  &  1 - \lambda_3 + \lambda_4  &  \lambda_1 + \lambda_2  &  0   \\
     0  &   \lambda_1 + \lambda_2  &  1 - \lambda_3 + \lambda_4  &  0  \\
      \lambda_1 - \lambda_2  &  0  &  0  &  \lambda_1 + \lambda_2
                     \end{array}                 \right).
\end{eqnarray} 
This is a boundary state in the convex set of the separable states, because the minimal 
eigenvalue of its partial transposition, {\it say} $\tilde{\sigma}^{\Gamma}$, is zero. Thus, the procedure
mentioned in the Introduction is terminated with identifying $\sigma_* = \tilde{\sigma}$. 
In fact, it is easy to show that $\tilde{\sigma}$ is exactly 
the same with $\pi_{BD}$ in Eq. (\ref{BD-3}). Thus, the procedure we suggested  correctly derives the CSS of the Bell-diagonal states.

\section{Generalized Vedral-Plenio State}
In this section we will derive the CSS of the generalized Vedral-Plenio (GVP) state defined as 
\begin{equation}
\label{VP-1}
\rho_{vp} = \lambda_1 \ket{\beta_3}\bra{\beta_3} + \lambda_2 \ket{01}\bra{01} + \lambda_3 \ket{10}\bra{10}
\hspace{1.0cm} (\lambda_1 + \lambda_2 + \lambda_3 = 1)
\end{equation}
by following the procedure mentioned above. In fact the CSS and REE of the GVP were 
explicitly derived in Ref.\cite{difficulty} using a geometric argument, which are 
\begin{eqnarray}
\label{VP-2}
& &\pi_{vp} = \left(\frac{\lambda_1}{2} + \lambda_2\right) \ket{01}\bra{01} + 
              \left(\frac{\lambda_1}{2} + \lambda_3\right) \ket{10}\bra{10}	             \\  \nonumber
& &{\cal E}_R (\rho_{vp}) = h \left( \frac{\lambda_1}{2} + \lambda_2 \right) - h (\Lambda_+)  
\end{eqnarray}
where
\begin{equation}
\label{VP-3}
 \Lambda_{\pm} = \frac{1}{2} \left[ 1 \pm \sqrt{\lambda_1^2 + 
    \left(\lambda_2 - \lambda_3\right)^2} \right].
\end{equation}

Now, we define 
\begin{equation}
\label{VP-4}
a = \frac{\lambda_1 \Lambda_+}{\sqrt{\lambda_1^2 + (\lambda_2 - \lambda_3)^2}}
\hspace{.5cm}
b = -\frac{\left( \lambda_2 - \lambda_3 \right) \sqrt{\Lambda_+ \Lambda_-}}
          {\sqrt{\lambda_1^2 + (\lambda_2 - \lambda_3)^2}}
\hspace{.5cm}
c = - \frac{\lambda_1 \Lambda_-}{\sqrt{\lambda_1^2 + (\lambda_2 - \lambda_3)^2}}
\end{equation}
and $\Omega^2 = 2[(a-c)^2 + 4 b^2 - (a-c) \sqrt{(a - c)^2 + 4 b^2}]$. We also define the unnormalized 
states $\ket{v_{\pm}} = \sqrt{\Lambda_{\pm}} \ket{\Lambda_{\pm}}$, where $\ket{\Lambda_{\pm}}$ are eigenstates
of $\rho_{vp}$; 
\begin{eqnarray}
\label{VP-5}
& &\ket{\Lambda_{+}} = \frac{1}{N} \left[\left( \sqrt{\lambda_1^2 + \left(\lambda_2 - \lambda_3\right)^2} + 
(\lambda_2 - \lambda_3) \right) \ket{01} + \lambda_1 \ket{10} \right]    \\   \nonumber
& &\ket{\Lambda_{-}} = \frac{1}{N} \left[\lambda_1 \ket{01} - \left( \sqrt{\lambda_1^2 + \left(\lambda_2 -
 \lambda_3\right)^2} + (\lambda_2 - \lambda_3) \right) \ket{10} \right].
\end{eqnarray}
In Eq. (\ref{VP-5}) $N$ is a normalization constant given by 
\begin{equation}
\label{VP-6}
N^2 = 2 \sqrt{\lambda_1^2 + \left(\lambda_2 - \lambda_3\right)^2} \left\{ \sqrt{\lambda_1^2 + \left(\lambda_2 - \lambda_3\right)^2} + (\lambda_2 - \lambda_3)      \right\}.
\end{equation}
Then, following Ref.\cite{woot-98}, the optimal decomposition of $\rho_{vp}$ for EOF is 
$\rho_{vp} = \sum_{j=1}^2 p_j \ket{\psi_j^{VP}} \bra{\psi_j^{VP}}$, where $p_1 = p_2 = 1/2$ and 
\begin{eqnarray}
\label{VP-7}
& &\ket{\psi_1^{VP}} = \frac{-i}{\Omega} \left[ 2b - i \left\{\sqrt{(a-c)^2 + 4 b^2} - (a-c) \right\} \right]
                               \left( \ket{v_+} + i \ket{v_-} \right)                       \\   \nonumber
& &\ket{\psi_2^{VP}} = \frac{-i}{\Omega} \left[ 2b + i \left\{\sqrt{(a-c)^2 + 4 b^2} - (a-c) \right\} \right]
                               \left( \ket{v_+} - i \ket{v_-} \right).
\end{eqnarray}

Following appendix A one can derive the CSS for $\ket{\psi_j^{VP}}$ directly. Then, one can realize that 
$\ket{\psi_1^{VP}}$ and $\ket{\psi_2^{VP}}$ have the same CSS, which is identical with $\pi_{vp}$. Thus, the
procedure also gives a correct CSS for the GVP states.

\section{Generalized Horodecki states}
In this section we will show that the procedure also generates the correct CSS for the generalized Horodecki 
states
\begin{equation}
\label{GH-1}
\rho_{H} = \lambda_1 \ket{\beta_3}\bra{\beta_3} + \lambda_2 \ket{00}\bra{00} + \lambda_3 \ket{11}\bra{11}
\end{equation}
with $\lambda_1 + \lambda_2 + \lambda_3 = 1$ and $\lambda_1 \geq 2 \sqrt{\lambda_2 \lambda_3}$\footnote{If 
$\lambda_1 \leq 2 \sqrt{\lambda_2 \lambda_3}$, $\rho_{H}$ becomes a separable state.}. The CSS and REE of 
$\rho_H$ were derived in Ref.\cite{difficulty} using a geometrical argument and the results are 
\begin{eqnarray}
\label{GH-2}
& &\pi_{H} = \frac{(\lambda_1 + 2 \lambda_2) (\lambda_1 + 2 \lambda_3)}{2} \ket{\beta_3}\bra{\beta_3} + 
             \frac{(\lambda_1 + 2 \lambda_2)^2}{4} \ket{00}\bra{00} + 
             \frac{(\lambda_1 + 2 \lambda_3)^2}{4} \ket{11}\bra{11}	
                                                                                    \nonumber  \\
& &{\cal E}_R (\rho_H) = \lambda_1 \ln \lambda_1 + \lambda_2 \ln \lambda_2 + \lambda_3 \ln \lambda_3
                   + 2 h \left( \frac{\lambda_1}{2} + \lambda_2 \right) - \lambda_1 \ln 2.
\end{eqnarray}

Following Ref.\cite{woot-98} one can straightforwardly construct the optimal decomposition of $\rho_H$ for 
EOF, which is $\rho_H = \sum_{j=1}^3 p_j \ket{\psi_j^{H}} \bra{\psi_j^{H}}$, where $p_1=p_2=p_3 = 1/3$ and 
\begin{eqnarray}
\label{GH-3}
& &\ket{\psi_1^{H}} = \sqrt{\lambda_1} \ket{\beta_3} + \sqrt{\lambda_2} \ket{00} + \sqrt{\lambda_3} \ket{11}
                                                                             \nonumber   \\
& &\ket{\psi_2^{H}} = \sqrt{\lambda_1} \ket{\beta_3} + \sqrt{\lambda_2} e^{i 2 \pi / 3}\ket{00} +
                      \sqrt{\lambda_3}  e^{-i 2 \pi / 3} \ket{11}                         \\  \nonumber
& &\ket{\psi_3^{H}} = \sqrt{\lambda_1} \ket{\beta_3} + \sqrt{\lambda_2} e^{i 4 \pi / 3}\ket{00} +
                      \sqrt{\lambda_3}  e^{-i 4 \pi / 3} \ket{11}.
\end{eqnarray}

In order to treat $\ket{\psi_j^{H}}$ as an unified manner let us consider $\ket{\phi} = \sqrt{\lambda_1}
\ket{\beta_3} + \sqrt{\lambda_2} e^{i\theta} \ket{00} + \sqrt{\lambda_3} e^{-i \theta}\ket{11}$. Then, 
$\lambda_{\pm}$ defined in Eq. (\ref{def-2-2}) is
\begin{equation}
\label{GH-4}
\lambda_{\pm} = \left(\frac{R \pm \left(\sqrt{\lambda_2} + \sqrt{\lambda_3} \right)}{2} \right)^2
\end{equation}
where $R = \sqrt{2 \lambda_1 + \left(\sqrt{\lambda_2} - \sqrt{\lambda_3} \right)^2}$. Since $\lambda_{\pm}$
is independent of $\theta$, this fact indicates that $\lambda_{\pm}$ of $\ket{\psi_j^{H}}$ are 
equal to Eq.(\ref{GH-4}) for all $j$. Following appendix A, it is straightforward to show that 
the Schmidt bases of $\ket{\phi}$ are
\begin{eqnarray}
\label{GH-5}
& &\ket{0_A} = \sqrt{\frac{\lambda_1}{R \left[R -\left(\sqrt{\lambda_2} - \sqrt{\lambda_3} \right) \right]}}
                                                                \ket{0} + 
           \sqrt{\frac{R - \left(\sqrt{\lambda_2} - \sqrt{\lambda_3} \right)}{2 R}} e^{-i \theta} \ket{1}
                                                                                   \nonumber  \\  
& &\ket{1_A} = -\sqrt{\frac{\lambda_1}{R \left[R +\left(\sqrt{\lambda_2} - \sqrt{\lambda_3} \right) \right]}}
                                                                \ket{0} + 
           \sqrt{\frac{R + \left(\sqrt{\lambda_2} - \sqrt{\lambda_3} \right)}{2 R}} e^{-i \theta} \ket{1}
                                                                                              \\  \nonumber
& & \hspace{3.0cm}
\ket{0_B} = e^{i \theta} \ket{0_A}   \hspace{1.0cm} \ket{1_B} = -e^{i \theta} \ket{1_A}.
\end{eqnarray}
Then the CSS $\sigma_{\phi}$ of $\ket{\phi}$ is 
\begin{eqnarray}
\label{GH-6}
\sigma_{\phi} &\equiv& \lambda_+ \ket{0_A 0_B} \bra{0_A 0_B} + \lambda_-  \ket{1_A 1_B} \bra{1_A 1_B}
                                                                          \nonumber  \\
&=& \left(             \begin{array}{cccc}
\frac{\lambda_1 + 2 \lambda_2}{2} - \frac{\lambda_1}{2 R^2} & {\cal A} e^{i \theta} &  {\cal A} e^{i \theta} &
\frac{\lambda_1}{2 R^2} e^{2 i \theta}                                                      \\
{\cal A} e^{-i \theta} &  \frac{\lambda_1}{2 R^2} &  \frac{\lambda_1}{2 R^2} & {\cal B} e^{i \theta}       \\
{\cal A} e^{-i \theta} &  \frac{\lambda_1}{2 R^2} &  \frac{\lambda_1}{2 R^2} & {\cal B} e^{i \theta}       \\
\frac{\lambda_1}{2 R^2} e^{-2 i \theta} & {\cal B} e^{-i \theta} & {\cal B} e^{-i \theta} & 
\frac{\lambda_1 + 2 \lambda_3}{2} - \frac{\lambda_1}{2 R^2}
                       \end{array}                 \right)
\end{eqnarray}
where 
\begin{eqnarray}
\label{GH-7}
& &{\cal A} = \frac{\sqrt{2 \lambda_1}}{4 R^2} \left[2 \sqrt{\lambda_2} + 
   \left(\sqrt{\lambda_2} + \sqrt{\lambda_3} \right)\left(\lambda_1 - 2 \sqrt{\lambda_2 \lambda_3}\right)
                                                \right]               \\   \nonumber
& &{\cal B} = \frac{\sqrt{2 \lambda_1}}{4 R^2} \left[2 \sqrt{\lambda_3} + 
   \left(\sqrt{\lambda_2} + \sqrt{\lambda_3} \right)\left(\lambda_1 - 2 \sqrt{\lambda_2 \lambda_3}\right)
                                                \right].
\end{eqnarray}
Thus, the CSS $\sigma_j$ of $\ket{\psi_j^{H}}$ can be obtained by letting $\theta = 0$, $2 \pi/3$, $4 \pi / 3$, respectively. 

Then, $\tilde{\sigma} = \sum_{j=1}^3 p_j \sigma_j$ with $p_j = 1/3 \hspace{.2cm}(j=1, 2, 3)$ reduces
\begin{eqnarray}
\label{GH-8}
\tilde{\sigma} = 
\left(                    \begin{array}{cccc}
       \frac{\lambda_1 + 2 \lambda_2}{2} - \frac{\lambda_1}{2 R^2} & 0 & 0 & 0       \\
       0 & \frac{\lambda_1}{2 R^2} & \frac{\lambda_1}{2 R^2} & 0                     \\
        0 & \frac{\lambda_1}{2 R^2} & \frac{\lambda_1}{2 R^2} & 0                     \\
        0 & 0 & 0 &  \frac{\lambda_1 + 2 \lambda_3}{2} - \frac{\lambda_1}{2 R^2}
                           \end{array}           \right).
\end{eqnarray}
However, $\tilde{\sigma}$ is not a boundary state in the convex set of the separable states, because the 
minimum eigenvalue of $\tilde{\sigma}^{\Gamma}$ is positive. Thus, we define
\begin{equation}
\label{GH-9}
\sigma_{*} = x \tilde{\sigma} + (1 - x) \rho_H. \hspace{.5cm} (0 \leq x \leq 1)
\end{equation}
The condition that the minimum eigenvalue of $\sigma_*^{\Gamma}$ is zero fixes $x$ as 
\begin{equation}
\label{GH-10}
x = \frac{R^2}{2 \lambda_1} \left(\lambda_1 + 2 \sqrt{\lambda_2 \lambda_3}\right).
\end{equation}
Inserting Eq.(\ref{GH-10}) into $\sigma_*$, one can show that $\sigma_*$ reduces to $\pi_H$. Thus, our 
procedure gives a correct CSS for the generalized Horodecki states.

\section{Less Symmetric States}
In the previous sections we have shown that the procedure generates the correct CSS and REE for various 
symmetric states such as Bell-diagonal, GVP, and generalized Horodecki states. In this section we will 
apply the procedure to the less symmetric states. 

\subsection{Vedral-Plenio-Type State}
The first quantum state we consider is 
\begin{equation}
\label{VPT-1}
\Sigma_1 = A_2 \ket{01} \bra{01} + A_3 \ket{10} \bra{10} + D \left( \ket{01}\bra{10} + \ket{10} \bra{01} \right),
\end{equation}
where $A_2 + A_3 = 1, A_2 \geq A_3$ and $0 \leq D \leq \sqrt{A_2 A_3}$. Of course, if 
$A_2 = \frac{\lambda_1}{2} + \lambda_2$, $A_3 = \frac{\lambda_1}{2} + \lambda_3$, and 
$D = \frac{\lambda_1}{2}$, $\Sigma_1$ reduces to $\rho_{vp}$ in Eq. (\ref{VP-1}). 
Thus, we call $\Sigma_1$ as Vedral-Plenio-type state.

In order to apply the procedure to $\Sigma_1$ we introduce 
\begin{eqnarray}
\label{VPT-2}
& &R = \sqrt{(A_2 - A_3)^2 + 4 D^2}  \hspace{1cm} \tan 2\theta = \frac{2 D}{A_2 - A_3}   \nonumber \\
& &\lambda_1 = \frac{1}{2} \left[\left( A_2 + A_3\right) + R \right]  
\hspace{1.0cm} \lambda_2 = \frac{1}{2} \left[\left( A_2 + A_3\right) - R \right]        \\   \nonumber
& &\ket{\lambda_1} = \cos \theta \ket{01} + \sin \theta \ket{10}
\hspace{1.0cm} \ket{\lambda_2} = \sin \theta \ket{01} - \cos \theta \ket{10}.
\end{eqnarray}
Applying Ref.\cite{woot-98}, it is possible to derive the optimal decomposition of $\Sigma_1$ for EOF;
$\Sigma_1 = p_1 \ket{w_1} \bra{w_1} + p_2 \ket{w_2} \bra{w_2}$, where
\begin{equation}
\label{VPT-3}
p_1 = \frac{1}{2} \left[1 + \frac{A_2 - A_3}{\sqrt{1 - 4 D^2}} \right]    \hspace{1.0cm}
p_2 = \frac{1}{2} \left[1 - \frac{A_2 - A_3}{\sqrt{1 - 4 D^2}} \right]
\end{equation}
and 
\begin{eqnarray}
\label{VPT-4}
& &\ket{w_1} = \frac{1}{{\mathcal Y}_+} \left[ \left(\sqrt{\xi_+ \eta_+} + \sqrt{\xi_- \eta_-} \right)
               \sqrt{\lambda_1} \ket{\lambda_1} + \left(\sqrt{\xi_+ \eta_-} - \sqrt{\xi_- \eta_+} \right)
               \sqrt{\lambda_2} \ket{\lambda_2} \right]     \\   \nonumber
& &\ket{w_2} = \frac{1}{{\mathcal Y}_-} \left[ \left(\sqrt{\xi_+ \eta_-} - \sqrt{\xi_- \eta_+} \right)
               \sqrt{\lambda_1} \ket{\lambda_1} - \left(\sqrt{\xi_+ \eta_+} + \sqrt{\xi_- \eta_-} \right)
               \sqrt{\lambda_2} \ket{\lambda_2} \right].
\end{eqnarray}
In Eq. (\ref{VPT-4}) $\xi_{\pm}$, $\eta_{\pm}$, and ${\mathcal Y}_{\pm}$ are 
\begin{eqnarray}
\label{VPT-5}
& &\xi_{\pm} = R \sqrt{A_2 A_3} \pm D \left(A_2 + A_3 \right)            \hspace{.5cm}
\eta_{\pm} = \sqrt{A_2 A_3 (1 - 4 D^2)} \pm D \left(A_2 - A_3 \right)            \\   \nonumber
& & \hspace{3cm} {\mathcal Y}_{\pm}^2 = 2 A_2 A_3 R \left[\sqrt{1 - 4 D^2} \pm \left(A_2 - A_3 \right) \right].
\end{eqnarray}

Following appendix A, one can derive the CSS $\sigma_1$ and $\sigma_2$ of $\ket{w_1}$ and $\ket{w_2}$ 
after long and tedious calculation. The final results are 
\begin{eqnarray}
\label{VPT-6}
\sigma_1&=& \left[\frac{\cos \theta \sqrt{\lambda_1} \left(\sqrt{\xi_+ \eta_+} + \sqrt{\xi_- \eta_-} \right) + 
\sin \theta \sqrt{\lambda_2} \left(\sqrt{\xi_+ \eta_-} - \sqrt{\xi_- \eta_+} \right) }{{\mathcal Y}_+}\right]^2
\ket{01} \bra{01}                                           \nonumber  \\
&+&\left[\frac{\sin \theta \sqrt{\lambda_1} \left(\sqrt{\xi_+ \eta_+} + \sqrt{\xi_- \eta_-} \right) - 
\cos \theta \sqrt{\lambda_2} \left(\sqrt{\xi_+ \eta_-} - \sqrt{\xi_- \eta_+} \right) }{{\mathcal Y}_+}\right]^2
\ket{10} \bra{10}                                                         \\  \nonumber
\sigma_2&=& \left[\frac{\cos \theta \sqrt{\lambda_1} \left(\sqrt{\xi_+ \eta_-} - \sqrt{\xi_- \eta_+} \right) - 
\sin \theta \sqrt{\lambda_2} \left(\sqrt{\xi_+ \eta_+} + \sqrt{\xi_- \eta_-} \right) }{{\mathcal Y}_-}\right]^2
\ket{01} \bra{01}                                             \\  \nonumber
&+&\left[\frac{\sin \theta \sqrt{\lambda_1} \left(\sqrt{\xi_+ \eta_-} - \sqrt{\xi_- \eta_+} \right) + 
\cos \theta \sqrt{\lambda_2} \left(\sqrt{\xi_+ \eta_+} + \sqrt{\xi_- \eta_-} \right) }{{\mathcal Y}_-}\right]^2
\ket{10} \bra{10}.
\end{eqnarray}
Then, $\tilde{\sigma} = p_1 \sigma_1 + p_2 \sigma_2$ simply reduces to 
\begin{equation}
\label{VPT-7}
\tilde{\sigma} = A_2 \ket{01} \bra{01} + A_3 \ket{10} \ket{10}.
\end{equation}
This is manifestly boundary state in the convex set of separable states. Thus, the procedure states that 
$\tilde{\sigma}$ is a CSS of $\Sigma_1$. This is exactly the same with theorem $1$ of Ref.\cite{z-axis}. 

\subsection{Horodecki-Type State}
The second less symmetric quantum state we consider is 
\begin{eqnarray}
\label{HT-1}
\Sigma_2 =  \left(                \begin{array}{cccc}
                  A_1  &  0  &  0  &  0                 \\
                  0  &  A  &  D  &  0                   \\
                  0  &  D  &  A  &  0                   \\
                  0  &  0  &  0  &  A_4
                                  \end{array}                   \right)
\end{eqnarray}
where $A_1 + A_4 + 2 A = 1$ and $\sqrt{A_1 A_4} < D \leq A$. If $A=D = \lambda_1 / 2$, $A_1 = \lambda_2$,
and $A_4 = \lambda_3$, $\Sigma_2$ reduces to $\rho_{H}$ in Eq. (\ref{GH-1}). Thus, we call $\Sigma_2$ as
Horodecki-type state. Applying Ref.\cite{woot-98}, one can derive the optimal decomposition of 
$\Sigma_2$ for EOF as $\Sigma_2 = \sum_{j=1}^4 p_j \ket{h_j}$, where $p_j = 1/4$ for all $j$ and
\begin{eqnarray}
\label{HT-2}
& &\ket{h_1} = \sqrt{A+D} \ket{\beta_3} + \sqrt{A-D} \ket{\beta_4} + \sqrt{A_1} \ket{00} + \sqrt{A_4} \ket{11}
                                                                                              \nonumber  \\
& &\ket{h_2} = \sqrt{A+D} \ket{\beta_3} + \sqrt{A-D} \ket{\beta_4} - \sqrt{A_1} \ket{00} - \sqrt{A_4} \ket{11}
                                                                                           \\   \nonumber
& &\ket{h_3} = \sqrt{A+D} \ket{\beta_3} - \sqrt{A-D} \ket{\beta_4} + i \sqrt{A_1} \ket{00} - 
                i \sqrt{A_4} \ket{11}                                                      \\   \nonumber
& &\ket{h_4} = \sqrt{A+D} \ket{\beta_3} - \sqrt{A-D} \ket{\beta_4} - i \sqrt{A_1} \ket{00} + 
                i \sqrt{A_4} \ket{11}.
\end{eqnarray}

In order to consider $\ket{h_j} \hspace{.2cm} (j=1,\cdots, 4)$ all together, we define
\begin{eqnarray}
\label{HT-3}
& &\ket{\varphi_1} = \sqrt{A + D} \ket{\beta_3} + \sqrt{A - D} \ket{\beta_4} + e^{i \theta} \sqrt{A_1}
                \ket{00} + e^{-i \theta} \sqrt{A_4} \ket{11}                 \\   \nonumber
& &\ket{\varphi_2} = \sqrt{A + D} \ket{\beta_3} - \sqrt{A - D} \ket{\beta_4} + e^{i \theta} \sqrt{A_1}
                \ket{00} + e^{-i \theta} \sqrt{A_4} \ket{11}.
\end{eqnarray}
For $\ket{\varphi_1}$ the Schmidt bases are 
\begin{eqnarray}
\label{HT-4}
& &\ket{0_A} = \frac{1}{2 {\mathcal Z}_+} 
\bigg[ \sqrt{2} \left( \sqrt{A - D} \sqrt{1 + {\mathcal C}} + \sqrt{A + D} \sqrt{1 - {\mathcal C}} \right)
                                                                                 \ket{0}   
                                                                                        \nonumber   \\
& &  \hspace{2.0cm}
+ e^{-i \theta} \left\{ \left(\sqrt{A_1} + \sqrt{A_4} \right) \sqrt{1 + {\mathcal C}} - 
                        \left(\sqrt{A_1} - \sqrt{A_4} \right)  \sqrt{1 - {\mathcal C}} \right\} \ket{1}
                                                                                      \bigg] 
                                                                                         \nonumber   \\
& &\ket{1_A} = \frac{1}{2 {\mathcal Z}_-} 
\bigg[ \sqrt{2} \left( \sqrt{A - D} \sqrt{1 + {\mathcal C}} - \sqrt{A + D} \sqrt{1 - {\mathcal C}} \right)
                                                                                 \ket{0}   
                                                                                         \\ \nonumber   
& &  \hspace{2.0cm}
+ e^{-i \theta} \left\{ \left(\sqrt{A_1} + \sqrt{A_4} \right) \sqrt{1 + {\mathcal C}} + 
                        \left(\sqrt{A_1} - \sqrt{A_4} \right)  \sqrt{1 - {\mathcal C}} \right\} \ket{1}
                                                                                      \bigg] 
                                                                                         \\  \nonumber  
& &\ket{0_B} = \frac{1}{2 {\mathcal Z}_+} 
\bigg[ \sqrt{2} e^{i \theta}\left\{ \sqrt{A + D} \left(\sqrt{A_1} + \sqrt{A_4} \right) + \sqrt{A - D} 
                                         \left(\sqrt{A_1} - \sqrt{A_4} \right)  \right\}
                                                                                 \ket{0}   
                                                                                          \\ \nonumber   
& &  \hspace{4.0cm}
     + \left\{ - \left(A_1 - A_4 \right) + 2 \sqrt{A^2 - D^2} + \sqrt{1 - {\mathcal C}^2} \right\}
                                                                               \ket{1}\bigg]
                                                                                             \\    \nonumber
& &\ket{1_B} = \frac{1}{2 {\mathcal Z}_-} 
\bigg[ \sqrt{2} e^{i \theta}\left\{ \sqrt{A + D} \left(\sqrt{A_1} + \sqrt{A_4} \right) + \sqrt{A - D} 
                                         \left(\sqrt{A_1} - \sqrt{A_4} \right)  \right\}
                                                                                 \ket{0}   
                                                                                          \\ \nonumber   
& &  \hspace{4.0cm}
     + \left\{ - \left(A_1 - A_4 \right) + 2 \sqrt{A^2 - D^2} - \sqrt{1 - {\mathcal C}^2} \right\}
                                                                               \ket{1}\bigg],
\end{eqnarray}
where ${\mathcal C} = 2 \left(D - \sqrt{A_1 A_4} \right)$ and  
\begin{equation}
\label{HT-5}
{\mathcal Z}_{\pm}^2 = \frac{1}{2} \sqrt{1 - {\mathcal C}^2}
    \left[  \sqrt{1 - {\mathcal C}^2} \mp \left(A_1 - A_4 \right) \pm 2 \sqrt{A^2 - D^2} \right].
\end{equation}
Thus, the CSS $\sigma_1 (\theta) $of $\ket{\varphi_1}$ is 
\begin{equation}
\label{HT-6}
\sigma_1 (\theta) = \left(\frac{\sqrt{1 + {\mathcal C}} + \sqrt{1 - {\mathcal C}}}{2} \right)^2
                    \ket{0_A 0_B}  \bra{0_A 0_B} + 
                    \left(\frac{\sqrt{1 + {\mathcal C}} - \sqrt{1 - {\mathcal C}}}{2} \right)^2
                    \ket{1_A 1_B}  \bra{1_A 1_B}.
\end{equation}
Similarly, it is straightforward to derive the CSS $\sigma_2 (\theta)$ of $\ket{\varphi_2}$.
Then, one can show 
\begin{eqnarray}
\label{HT-7}
\tilde{\Pi} &\equiv& \frac{1}{4} \left[\sigma_1 (0) + \sigma_1 (\pi) + \sigma_2 \left(\frac{\pi}{2} \right) + 
                         \sigma_2 \left(-\frac{\pi}{2} \right)  \right]                
                                                                          \nonumber  \\
& &   \hspace{2.0cm} =    \left(                  \begin{array}{cccc}
                a_1  &  0  &  0  &  0                    \\
                0  &  a  &  d  &  0                      \\
                0  &  d  &  a  &  0                      \\
                0  &  0  &  0  &  a_4
                                 \end{array}                  \right)
\end{eqnarray}
where 
\begin{eqnarray}
\label{HT-8}
& &a_1 = \frac{1}{4 (1 - {\mathcal C}^2)} 
\bigg[ (1 + {\mathcal C}) \left(\sqrt{A_1} + \sqrt{A_4} \right)^2 +  (1 - {\mathcal C}) 
\left(\sqrt{A_1} - \sqrt{A_4} \right)^2                                       \nonumber   \\
& &  \hspace{9.0cm}
                + 2  (1 - {\mathcal C}^2) \left(A_1 - A_4 \right) \bigg]       \nonumber   \\
& &a_4 = \frac{1}{4 (1 - {\mathcal C}^2)} 
\bigg[ (1 + {\mathcal C}) \left(\sqrt{A_1} + \sqrt{A_4} \right)^2 +  (1 - {\mathcal C}) 
\left(\sqrt{A_1} - \sqrt{A_4} \right)^2                                       \nonumber   \\
& &  \hspace{9.0cm}
                - 2  (1 - {\mathcal C}^2) \left(A_1 - A_4 \right) \bigg]     \\  \nonumber  
& &a = \frac{1}{2 (1 - {\mathcal C}^2)} \left[ (1 + {\mathcal C}) \left(A - D \right) + 
                                               (1 - {\mathcal C}) \left(A + D \right)   \right]
                                                                             \\   \nonumber
& & d = \frac{2 A \sqrt{A_1 A_4} + D \left( A_1 + A_4 \right)}{1 - {\mathcal C}^2}.
\end{eqnarray} 
One can show that if $A=D = \lambda_1 / 2$, $A_1 = \lambda_2$, and $A_4 = \lambda_3$, $\tilde{\Pi}$ reduces to 
Eq. (\ref{GH-8}). 

Since $\tilde{\Pi}$ is not a boundary state in the set of separable states, we define
\begin{equation}
\label{HT-9}
\Pi_* = x \tilde{\Pi} + (1 - x) \Sigma_2.
\end{equation}
Then, the CSS condition of $\Pi_*$ is 
\begin{equation}
\label{HT-10}
\left[x \left(a_1 - A_1 \right) + A_1 \right] \left[x \left(a_4 - A_4 \right) + A_4 \right]
= \left[x (d - D) + D \right]^2.
\end{equation}
In the Horodecki state limit Eq. (\ref{HT-10}) gives a solution (\ref{GH-10}).
Using $a_1 - A_1 = a_4 - A_4 = - (a - A) = f / (1 - {\mathcal C}^2)$ and $d - D = g / (1 - {\mathcal C}^2)$
where
\begin{equation}
\label{HT-11}
f = {\mathcal C} \left(D - A {\mathcal C} \right)     \hspace{1.0cm}
g = {\mathcal C} \left( {\mathcal C} D - A \right),
\end{equation}
the solution of $x$, {\it say} $x = x_*$, can be obtained by solving the quadratic 
equation  (\ref{HT-10}). Inserting $x = x_*$ in Eq. (\ref{HT-9}), one can compute $\Pi_*$ explicitly, which is 
a candidate of CSS for $\Sigma_2$. 

The CSS of $\Sigma_2$ was derived in the theorem $2$ of Ref.\cite{z-axis} by using the converse procedure
introduced in Ref.\cite{miran-08-1}. The explicit form of the CSS is 
\begin{eqnarray}
\label{HT-12}
\pi_{\Sigma_2} = \left(                \begin{array}{cccc}
                        r_1  &  0  &  0  &  0      \\
                        0  &  r  &  y  &  0        \\
                        0  &  y  &  r  &  0        \\
                        0  &  0  &  0  &  r_4
                                        \end{array}           \right)
\end{eqnarray} 
where
\begin{eqnarray}
\label{th-2-3}
& &r_1 = \frac{1}{F} \left[2A_1 (A_1 + A_2) (A_1 + A_2 + A_4) - D^2 (A_1 - A_4) + \Delta \right]
                                                                                                        \\   \nonumber
& &r_4 = \frac{1}{F} \left[2A_4 (A_2 + A_4) (A_1 + A_2 + A_4) + D^2 (A_1 - A_4) + \Delta \right]
                                                                                                        \\   \nonumber
& &r = \frac{1}{F} \left[ 2(A_1 + A_2) (A_2 + A_4) (A_1 + A_2 + A_4) - D^2 (A_1 + 2 A_2 + A_4) - \Delta \right]
\end{eqnarray}
and $ y = \sqrt{r_1 r_4}$. In Eq. (\ref{th-2-3}) $D$ and $\Delta$ are
\begin{eqnarray} 
\label{th-2-4}
& & F = 2 (A_1 + A_2 + A_4 + D) (A_1 + A_2 + A_4 - D)                                \\  \nonumber
& & \Delta = D \sqrt{D^2 (A_1 - A_4)^2 + 4 A_1 A_4 (A_1 + A_2) (A_2 + A_4)}.
\end{eqnarray}
Our candidate $\Pi_*|_{x = x_*}$ does not coincide with the correct CSS $\pi_{\Sigma_2}$. Thus, the procedure 
does not give a correct REE for $\Sigma_2$, although it gives correct REE for Bell-diagonal, GVP, generalized
Horodecki, and Vedral-Plenio-type states.

\section{Conclusion}
In this paper we examine the possibility for deriving the closed formula for REE in two-qubit system
without relying on the converse procedure discussed in Ref.\cite{miran-08-1,converse-1,converse-2}. 
Since REE and EOF are identical for all pure states in spite of their different definitions, we think they 
should have some connection somehow. In this context we suggest a procedure, where REE can be computed
from EOF. The procedure gives correct REE for many symmetric states such as Bell-diagonal, GVP, and 
generalized Horodecki states.  It also generates a correct REE for less symmetric states such as $\Sigma_1$. 
However, the procedure failed to produce a correct REE for the less symmetric states $\Sigma_2$. 
This means our procedure is still incomplete for deriving the closed formula of REE. 

 We think still the connection between EOF and REE is not fully revealed. If this connection is 
 sufficiently understood in the future, probably the closed formula for REE can be derived.  We 
 hope to explore this issue in the future.

{\bf Acknowledgement}:
This research was supported by the Basic Science Research Program through the National Research Foundation of Korea(NRF) funded by the Ministry of Education, Science and Technology(2011-0011971).

\newpage

\begin{appendix}{\centerline{\bf Appendix A}}

\setcounter{equation}{0}
\renewcommand{\theequation}{A.\arabic{equation}}
In this section we will show that REE and EOF are identical for two-qubit pure states. This fact was already proven in Theorem $3$ of 
Ref.\cite{vedral-97-2}. We will prove this again more directly, 
because explicit Schmidt bases are used in the main body of the paper. 

Let us consider
a general two-qubit pure state $\ket{\psi_2}_{AB} = \alpha_1 \ket{00} + \alpha_2 \ket{01} + \alpha_3 \ket{10} + \alpha_4 \ket{11}$ with 
$|\alpha_1|^2 + |\alpha_2|^2 + |\alpha_3|^2 + |\alpha_4|^2 = 1$. Then, its concurrence is ${\cal C} = 2 |\alpha_1 \alpha_4 - \alpha_2 \alpha_3|$. 
Now, we define 
\begin{equation}
\label{def-2-1}
 x_{\pm} = \frac{\alpha_1^* \alpha_2 + \alpha_3^* \alpha_4}{{\cal N}_{\pm}}         \hspace{1.0cm}
y_{\pm} = \frac{\lambda_{\pm} - (|\alpha_1|^2 + |\alpha_3|^2)}{{\cal N}_{\pm}}        
\end{equation}
where 
\begin{equation}
\label{def-2-2}
\lambda_{\pm} = \frac{1}{2} \left[1 \pm \sqrt{1 - {\cal C}^2} \right]     \hspace{1.0cm}
{\cal N}_{\pm}^2 = |\alpha_1^* \alpha_2 + \alpha_3^* \alpha_4|^2 + |\lambda_{\pm} - (|\alpha_1|^2 + |\alpha_3|^2)|^2.
\end{equation}
Now, we consider $2 \times 2$ matrix $u$, whose components $u_{ij}$ are 
\begin{eqnarray}
\label{def-2-3}
& & u_{11} = \alpha_1 \left( \frac{|x_+|^2}{\sqrt{\lambda_+}} +  \frac{|x_-|^2}{\sqrt{\lambda_-}} \right) 
       + \alpha_2 \left( \frac{x_+^* y_+}{\sqrt{\lambda_+}} + \frac{x_-^* y_-}{\sqrt{\lambda_-}} \right)     \nonumber   \\
& &u_{12} = \alpha_1 \left( \frac{x_+ y_+^*}{\sqrt{\lambda_+}} +  \frac{x_- y_-^*}{\sqrt{\lambda_-}} \right) 
       + \alpha_2 \left( \frac{|y_+|^2}{\sqrt{\lambda_+}} + \frac{|y_-|^2}{\sqrt{\lambda_-}} \right)           \\    \nonumber
& &u_{21} = \alpha_3 \left( \frac{|x_+|^2}{\sqrt{\lambda_+}} +  \frac{|x_-|^2}{\sqrt{\lambda_-}} \right) 
       + \alpha_4 \left( \frac{x_+^* y_+}{\sqrt{\lambda_+}} + \frac{x_-^* y_-}{\sqrt{\lambda_-}} \right)        \nonumber  \\
& &u_{22} = \alpha_3 \left( \frac{x_+ y_+^*}{\sqrt{\lambda_+}} +  \frac{x_- y_-^*}{\sqrt{\lambda_-}} \right) 
       + \alpha_4 \left( \frac{|y_+|^2}{\sqrt{\lambda_+}} + \frac{|y_-|^2}{\sqrt{\lambda_-}} \right).           \nonumber
\end{eqnarray}
Then Schmidt bases for each party are defined as 
\begin{equation}
\label{schmidt-1}
\ket{i_A} = \sum_{j=0}^1 v_{ji} \ket{j}     \hspace{1.0cm} \ket{i_B} = \sum_{k=0}^1 w_{ik} \ket{k}   \hspace{.5cm} (i = 0, 1)
\end{equation}
where 
\begin{eqnarray}
\label{schmidt-2}
v = \left(    \begin{array}{cc}
               u_{11}  &  u_{12}   \\
               u_{21}  &  u_{22}
               \end{array}            \right)
    \left(    \begin{array}{cc}
               x_+  &  x_-   \\
               y_+  &  y_-
               \end{array}            \right)                \hspace{1.0cm}
w = \left(    \begin{array}{cc}
               x_+^*  &  y_+^*   \\
               x_-^*  &  y_-^*
               \end{array}            \right).
\end{eqnarray}
Using Eq. (\ref{schmidt-1}), one can show straightforwardly that $\ket{\psi_2}_{AB}$ reduces to 
$\ket{\psi_2}_{AB} = \sqrt{\lambda_+} \ket{0_A 0_B} + \sqrt{\lambda_-} \ket{1_A 1_B}$. Thus, its 
CSS $\sigma_*$ are simply expressed in terms of the Schmidt bases as 
\begin{equation}
\label{result-2}
\sigma_{*} = \lambda_+  \ket{0_A 0_B} \bra{0_A 0_B} + \lambda_- \ket{1_A 1_B} \bra{1_A 1_B}.
\end{equation}
Applying Eq. (\ref{ree-1-1}), one can show easily ${\cal E}_R (\ket{\psi_2}) = - \lambda_+ \ln \lambda_{+} - \lambda_- \ln \lambda_{-}$, which is 
exactly the same with EOF.

\end{appendix}

\end{document}